\documentclass[journal,twoside,web]{ieeecolor}
\usepackage{jsen}
\usepackage{cite}
\usepackage{amsmath,amssymb,amsfonts}
\usepackage{algorithmic}
\usepackage{graphicx}
\usepackage{textcomp}
\usepackage{wrapfig}
\usepackage{caption}
\usepackage{algorithm}
\usepackage{booktabs}

\def\BibTeX{{\rm B\kern-.05em{\sc i\kern-.025em b}\kern-.08em
    T\kern-.1667em\lower.7ex\hbox{E}\kern-.125emX}}
\definecolor{abstractbg}{rgb}{0.89804,0.94510,0.83137}
\begin{document}
\title{GeXSe (Generative Explanatory Sensor System): A  Deep Generative method for Human Activity Recognition of Smart Spaces IOT}
         	

\author{Sun Yuan, Salami Pargoo Navid,  Ortiz Jorge 
\thanks{Sun Yuan, ys820@soe.rutgers.edu, Salami Pargoo, Navid navid.salamipargoo@rutgers.edu, Walter, Hedaya Maryam hedaya.walter@rutgers.edu,  Ortiz Jorge jorge.ortiz@rutgers.edu }
\thanks{Rutgers, The State University of New Jersey}
\thanks{ }}

\IEEEtitleabstractindextext{%
\fcolorbox{abstractbg}{abstractbg}{%
\begin{minipage}{\textwidth}%


\begin{wrapfigure}[12]{r}{3in}%
\includegraphics[width=3in]{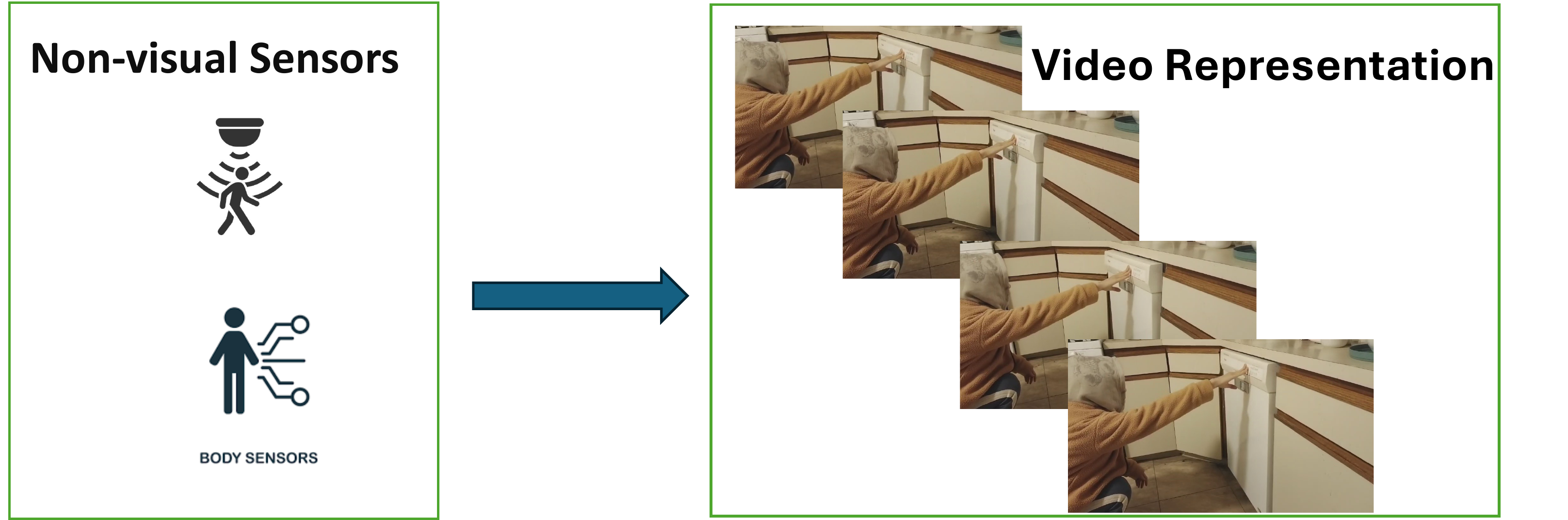}%
\end{wrapfigure}%

\begin{abstract}
Effective sensemaking is crucial across various domains where trust in system outputs is essential, highlighting a significant challenge in relying solely on sensor-based activity recognition. This limitation underscores the need for interpretable models that process raw data and provide comprehensible insights. Our work introduces GeXSe, a novel multi-task framework that jointly models raw sensors for classification while also generating grounded visual explanations. At the core of GeXSe lies a Parallel Multi-Branch Multi-Layer Perceptron Fast Fourier Convolution (PMB-MLP-FFC) module tailored for multi-modal sensor fusion and explanation generation. PMB-MLP-FFC extracts interpretable features optimized for both tasks through multi-branch parallel convolutions and Fourier transforms. We validated GeXSe across 3 diverse public datasets of daily activities recorded by camera, microphone, motion, and environmental sensors. Results show superior activity recognition over baseline models. Furthermore, human evaluation studies confirm the generated visual explanations enhance understanding and trust compared to purely sensor-based outputs. 
\end{abstract}


\end{minipage}  }}

\maketitle

\section{Introduction }
\label{chap:1}

The rapid advancement of deep learning techniques has revolutionized Human Activity Recognition (HAR), enabling highly accurate and complex models for interpreting human behaviors from sensor data. As these models become more sophisticated, there is a growing need for interpretability and explainability to ensure trust and transparency in HAR systems. Without understanding how these models make decisions and provide meaningful explanations, their adoption in critical applications, such as healthcare and smart environments, may be hindered.

Interpretability involves comprehending the internal mechanisms of a model and/or gaining insights into the conclusions drawn by that model.  It focuses on the transparency of the model's structure and decision-making process or result. On the other hand, explainability is concerned with understanding why a model made a specific prediction for a given input, attributing the output back to the relevant input features~\cite{varshney2018interpretability}. While interpretability deals with the overall logic of the model, explainability provides insights into individual predictions~\cite{marquessilva2023logicbased, DBLP:journals/corr/abs-2102-02671}. Both aspects are crucial in HAR applications, where trust in the system's outputs is paramount, and users need to comprehend the reasoning behind the model's decisions.

\textbf{Example of Interpretability} Consider a HAR system using a decision tree to classify human activities based on sensor data. Each node in the tree represents a decision based on input from accelerometers, gyroscopes, and environmental sensors. For instance, a node might split activities based on accelerometer variance, with lower variances indicating stationary activities and higher variances suggesting movement. Further nodes differentiate between specific activities using gyroscope and environmental sensor patterns and thresholds.

Interpretability is achieved through the transparency of the decision tree structure. Users can comprehend the model's internal mechanisms by following the decision paths for classifying an activity. For example, if the model classifies an activity as `running,' users can trace the decision tree to understand that this classification was based on high accelerometer variance, a specific gyroscope speed threshold, and environmental factors aligning with typical running conditions.

\textbf{Example of Explainability} Explainability is demonstrated when the HAR system provides a detailed explanation for a specific prediction. For instance, when the system classifies a segment of sensor data as `jogging,' it might provide the following explanation: ``This prediction was made because the accelerometer showed consistent, moderate variance indicative of repetitive motion, the gyroscope detected a steady forward movement at a jogging pace, and the environmental sensors detected an outdoor setting with moderate temperature and GPS data indicating a path through a park, which aligns with typical jogging behavior.''

This explanation attributes the prediction to relevant input features from the sensor data, providing a clear rationale that links the data characteristics to the model's conclusion. Explainability increases trust in the system's decision-making process by offering insights into why the model made a particular prediction. It allows users to verify the reasoning against their domain knowledge.

Traditional HAR approaches, such as rule-based systems and classical machine learning algorithms like decision trees and linear regression, offer inherent interpretability due to their simple and transparent structures. However, these methods often struggle to capture the complex patterns and relationships in high-dimensional sensor data~\cite{Pandey2023ACA}, which can be difficult to interpret due to the combination of multiple sensors, such as accelerometers, gyroscopes, and environmental sensors, each generating unique patterns and relationships~\cite{wolf2019explainability}. Distinguishing between ambiguous activities, such as ``sitting'' and ``standing,'' or ``jogging'' and ``running,'' requires the model to capture subtle differences in sensor data. While the field has witnessed a paradigm shift with the adoption of deep learning techniques, offering robust solutions to this challenge ~\cite{chen2012sensor,bulling2014tutorial,wang2019deep,10.1145/3472290, 8782290}, the black-box nature of these models makes them difficult to interpret and explain~\cite{arrieta2020explainable}, hindering their adoption in critical applications where accountability and user trust are essential~\cite{Parker2019DEMISeID, DBLP:journals/corr/abs-1907-06194, Arrotta2022KnowledgeIF, Chew2022EnhancingIO}.

Recent studies~\cite{Zhou_2023,wang2022point} introduced saliency maps and feature importance scores to address the interpretability challenge in deep-learning-based HAR. However, these approaches often generate abstract visualizations or numerical values that may not be intuitive or actionable for users without machine learning expertise~\cite{aquino2022explaining}. This lack of comprehensible explanations can lead to a disconnect between the model's outputs and the users' understanding, ultimately impacting the trust and acceptance of HAR systems. Furthermore, most research on sensor data interpretation has focused on generating text-based, semantic explanations~\cite{bayat2014study, ravi2016deep, penatti2017human, zheng2018comparison, arrotta2022dexar, jeyakumar2023x, khaertdinov2023explaining}, with a notable gap in the use of vision-based, pictorial representations~\cite{hur2018iss2image}.

\begin{figure}[t]
  \centering
  \includegraphics[width=1\linewidth]{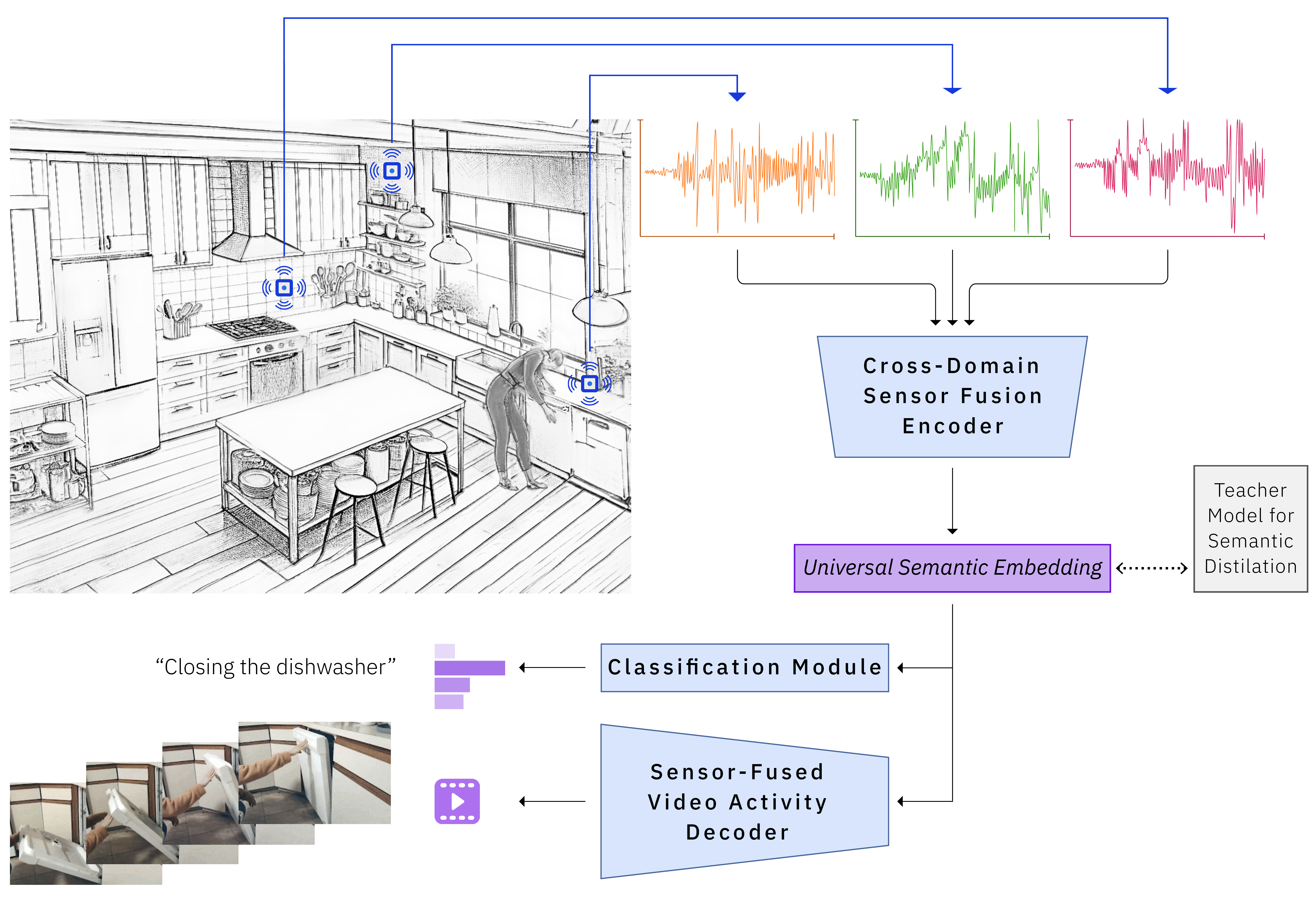}
  \vspace{0 em} 
  \captionsetup{justification=raggedright,singlelinecheck=false}
  \caption{\small{GeXSe Framework: This schematic represents a system for integrating diverse sensor inputs into a cohesive latent space using a Cross-Domain Sensor Fusion Encoder, which then interfaces with a Teacher Model to enhance the Universal Semantic Embedding. A Classification Module interprets the embedded signals, while a Sensor-Fused Video Activity Decoder translates the embeddings into annotated video activity, such as ``Closing the dishwasher,'' demonstrating the framework's capability for zero-shot video understanding by leveraging sensor-video correlations.}}
  \label{overallstructure}
\end{figure}

Deep generative models have emerged as a promising solution for deriving interpretable insights from complex sensor data. They bridge the gap between the technical complexity of sensor-based HAR and the provision of clear, actionable insights that resonate with human experiences.

\subsection{Knowledge Gap}
The growing interest in interpretable and explainable HAR models that rely on vision domain features is propelled by its potential to bolster the accuracy of activity recognition models~\cite{ha2016convolutional}. Despite various approaches~\cite{arrotta2022dexar, jeyakumar2023x, gochoo2017dcnn, li2016deep, tan2018multi}, the majority of current techniques for sensor-based HAR interpretations and explanations hinge on abstract visualizations, text descriptions, or labels, which may not always be intuitive or actionable for end-users. There is a need for explanation techniques that can effectively translate complex sensor data into \emph{visually grounded} and \emph{easily comprehensible} explanations, catering to end-users' needs in various domains\cite{hur2018iss2image,arrotta2022dexar,jeyakumar2023x}.

To address the challenge of translating complex sensor data into meaningful and understandable insights for end-users, a comprehensive and user-centric explainability framework is essential. This framework should focus on bridging the gap between the intricate patterns in sensor data and the user's domain knowledge, ensuring that the explanations provided by the HAR system are not only accurate but also intuitive, actionable, and aligned with the user's understanding of the activities and context. Such a framework would not only facilitate the extraction of features from sensor data but also align these features with vision domain descriptions, thereby enabling a deeper understanding of model outcomes~\cite{arrotta2022dexar,jeyakumar2023x,gochoo2017dcnn,li2016deep,tan2018multi}. This alignment is vital for creating explicative models that can offer explanations intelligible to humans, fostering a genuine sense of explainability.

Moreover, a critical challenge is designing a sensor model capable of effectively interpreting cross-domain sensor data with minimal complexity. Conventional methods require isolated training for each distinct dataset (domain) due to the inherent personalized characteristics of sensor data within smart environments~\cite{ribeiro2016should,lundberg2017unified,selvaraju2017grad}. Creating a universal sensor model capable of aligning semantic spaces is essential for translating sensor features into human-readable explanations. Additionally, developing a cost-effective strategy for transforming sensor data into vision domain features is crucial to ensure the approach is widely implementable and sustainable, enhancing the accessibility and feasibility of advanced sensor-based applications in smart environments.
\subsection{Contributions}

This paper introduces several key contributions in sensor data interpretation and translation into the vision domain, thereby enhancing the explainability and applicability of sensor-based activity detection systems. it is also the first paper address this problem in vision domain. Current method mostly focus on NLP based solution\cite{arrotta2022dexar,jeyakumar2023x}.Our main contributions are as follows:

\begin{itemize}
    \item \textbf{GeXSe Framework:} A novel framework\ref{overallstructure} for intuitively translating sensor data into vision domain features, intelligible to non-experts, enabling multimodal data representation.
    \item \textbf{Multi-Task Encoder:} A multi-task encoder coupled with the PMB-MLP-FFC model structure, showcasing high performance in smart space activity classification while generating universal representations across various sensor types, facilitating zero-shot downstream tasks by eliminating the need for labeling. Design its multitask loss helps to align the cross-domain sensor data into semantic representation.
    \item \textbf{Universal Decoder:} A pre-trained universal decoder based on stable diffusion for the downstream task of multimodal data representation, reducing computational demands and overall costs by translating semantic representations in a zero-shot fashion.
    \item \textbf{Symbolic Reasoning:} Implementation of symbolic reasoning to represent varying sensor data contributions as visual cues within video outputs across different activities and conditions, establishing a low-cost, efficient translation system without necessitating decoder retraining.
    \item \textbf{Cross-Domain Robustness:} A universal semantic representation enables the model adapting across diverse domains (various publicly available datasets), underscoring the effectiveness and versatility of GeXSe in real-world applications.

\end{itemize}

These contributions collectively advance the field by enhancing the interpretability and accessibility of sensor-based activity detection systems, facilitating intuitive and user-friendly applications in smart spaces.

\section{Related Work}
\label{chap:3}

\subsection{Activity Sensing in Smart Spaces}

\subsubsection{Camera-based Human Activity Recognition (HAR)}
Camera technology has been pivotal in advancing HAR within the domain of computer vision, supporting a wide array of applications~\cite{hu2004survey,erol2007vision,pavlovic1997visual,zhao2003face}. The declining cost of camera technology enhances its accessibility, though challenges in physical interface accessibility persist~\cite{laput2017synthetic,fails2003design}. 

The evolution towards \emph{general-purpose sensing} represents a shift from relying solely on camera-based methods to embracing interconnected, remotely controllable devices. This transition facilitates accessibility, mitigating the barriers associated with camera deployment, including privacy concerns and deployment costs~\cite{guo2016vizlens,boyle2000effects,beckmann2004some}. Consequently, the role of cameras in general-purpose sensing is reevaluated, acknowledging the limitations in capturing certain signals~\cite{raspberryshakelisten}.

\subsubsection{The Emergence of General-Purpose Sensing}

General-purpose sensing embodies the capability to monitor a variety of activities across smart environments through minimal sensor deployment. Modern sensor boards, equipped with a multitude of sensors, offer a versatile foundation for ubiquitous computing applications without necessitating hardware modifications~\cite{laput2017synthetic}. This approach fosters a distributed sensing paradigm, enhancing the scalability of activity detection and interaction within smart spaces. Unlike traditional sensing methods, general-purpose sensing leverages server-based machine learning analytics to process and interpret the raw data collected, promising enhancements in practical applications through deep learning and explainability~\cite{zhou2021internet}.

Building on the insights from previous research, it becomes evident that \emph{integrating the strengths of both vision-based and non-invasive sensing modalities remains underexplored}. To bridge this gap, we introduce GeXSe, a system designed to enrich video descriptions by synergistically combining camera-derived insights with data from non-invasive sensors. This approach allows us to leverage the detailed environmental context provided by vision systems alongside the discreet, continuous monitoring capabilities of non-invasive sensors, offering a comprehensive and enhanced depiction of activities within smart spaces.

While general-purpose sensing offers a versatile and scalable approach to activity detection and interaction within smart spaces, the complexity of the machine learning algorithms employed in these systems can hinder their interpretability and transparency. This lack of clarity in the decision-making process of AI-driven systems has led to the emergence of explainable AI (XAI), which aims to address these concerns.

\subsubsection{Explainable AI}
The complexity of advanced machine learning methods, particularly in analyzing non-linear signals, poses significant challenges for non-experts~\cite{shin2021effects}. This complexity hinders the adoption of deep learning tools in various domains, such as personalized dermatology, where non-specialists struggle to utilize these tools for diagnostic purposes~\cite{gomolin2020artificial}. To address these concerns, the field of explainable AI (XAI) has emerged, focusing on developing methods to enhance the interpretability and transparency of AI-driven systems~\cite{amgoud2009using,kohl2019explainability,langer2021we,paez2019pragmatic}.

To address these concerns, the field of explainable AI (XAI) has gained attention, focusing on making AI-driven systems more understandable~\cite{amgoud2009using,kohl2019explainability,langer2021we,paez2019pragmatic}. XAI methods aim to clarify how AI models make decisions and can be divided into two main categories: transparent models and post-hoc explanation techniques. Transparent models, such as linear regressions and rule-based learning, are easier to interpret but may struggle with complex data relationships~\cite{mood2010logistic,noh2012intelligent}. Post-hoc methods, on the other hand, provide explanations after the model has made a decision. These include 1) LIME (Local Interpretable Model-agnostic Explanations): A technique that creates simple, interpretable models to approximate the behavior of complex models locally around a specific input~\cite{ribeiro2016should}, 2) SHAP (SHapley Additive exPlanations): A method that assigns importance values to each input feature based on its contribution to the model's output~\cite{lundberg2017unified}, and 3) GradCAM (Gradient-weighted Class Activation Mapping): A technique that highlights the regions of an input image that are most important for a model's decision~\cite{selvaraju2017grad}.
While these post-hoc methods provide insights into model decisions, especially in computer vision, they primarily focus on visual explanations and may not capture complex feature interactions.

Notably, GradCAM's application extends to sensor data by converting such data into ``image'' matrices, adapting to its image-centric analysis framework~\cite{arrotta2022dexar}. This adaptation enables the utilization of GradCAM for multi-channel sensor inputs, aligning with its pixel-based processing capabilities. However, the reliance on semantic explanations underscores a broader challenge in delivering comprehensible visual domain explanations, highlighting the ongoing need for more accessible interpretative mechanisms in leveraging sensor data for actionable insights.

This study addresses the challenge of insufficient explanation methods for the visual domain of sensor data. Our objective is to devise an explanation technique finely tuned to sensor data. Our methodology introduces a video-based explanation for non-intrusive sensor data.

\subsubsection{Multi-task learning}

Multi-task learning (MTL) is a paradigm in machine learning that aims to improve the learning efficiency and prediction accuracy of a model by simultaneously learning multiple related tasks. By sharing representations between related tasks, MTL can leverage commonalities and differences across tasks to generalize better on each task than when learning tasks independently. This approach is particularly beneficial in scenarios where data for some tasks are limited, allowing these tasks to benefit from the knowledge acquired from data-rich tasks. MTL has been successfully applied in various domains, including natural language processing\cite{durrett2014joint}, computer vision\cite{kim2023cross}, and sensor data recognition\cite{ordonez2016deep}, demonstrating its versatility and effectiveness in enhancing model performance across multiple tasks.

We introduce a novel loss function tailored for multi-task learning, aiming to facilitate the model's ability to acquire a universal representation across diverse domains.

\subsection{Synthetic Vision Domain Description}

\subsubsection{Visual representations}
Visual output from sensor data offers several advantages. It captures non-verbal context and subtle details, providing a comprehensive representation of spatial relationships, postures, and interactions within an environment~\cite{hagmann2016ultrafast}. Moreover, visual representations can transcend language barriers and be universally understood, regardless of linguistic background~\cite{kveraga2014scene}.

\subsubsection{Diffusion model}

One of the most related booming areas is image generation via text.It generate synthesis image from text based image generation technique~\cite{goodfellow2020generative,mansimov2015generating,reed2016learning,brock2018large,karras2020analyzing,razavi2019generating}. The development of DALL-E2~\cite{dalle2_web}  has revolutionized text-to-image models~\cite{ramesh2021zero}  by introducing a new level of creativity and sophistication. Unlike previous models, which often produced limited or generic images based on textual descriptions, DALL-E2 has the ability to generate unique and imaginative visuals that go beyond what is explicitly stated in the text. This breakthrough has opened up new avenues for creative expression and has the potential to transform industries such as design, advertising, and entertainment. Diffusion model which improves these previous results via diffusion method that beat the state of the art result~\cite{dhariwal2021diffusion}. Not only texts, sensors that have more real-life "descriptions", also need to derive explainable vision domain features for humans to interact with. The representation of sensor data as an image has been a subject of increasing attention and significant focus~\cite{hur2018iss2image}. Various methods~\cite{tan2018multi,gochoo2017dcnn,li2016deep} have been attempted to convert sensor data into an "image," but none of them produce actual images. Instead, these methods are primarily based on natural language processing (NLP) and signal graph techniques.

In the realm of generative techniques, diffusion models have garnered significant attention due to their recent achievements. To begin with, a novel approach was presented involving the prediction of noise in image generation. This method produces images from raw Gaussian noises by iteratively refining the noise at each step. Building upon this concept, a plethora of enhancements have been suggested, primarily concentrating on enhancing the quality\cite{rombach2022high} of generated samples and optimizing the efficiency\cite{song2020denoising} of the sampling process. The diffusion method has been applied across various domains such as point clouds\cite{luo2021diffusion}, audio\cite{huang2023make}, and video\cite{ho2022imagen} generation.

All of these models utilizing diffusion techniques are constructed upon the foundation of the U-Net architecture, with particular emphasis on video generation, recognized as one of the most demanding tasks in the field of machine vision. The prevailing approach for video generation and prediction revolves predominantly around the utilization of the U-Net architecture\cite{ho2022imagen,blattmann2023align,voleti2022mcvd,yang2022diffusion}.

\subsubsection{Neural symbolic}

A pivotal aspect of consideration lies in the alignment of purely perception-driven models with the tenets of explainable AI\cite{ratti2022explainable}. Contrary to neural networks, which act as opaque entities incapable of revealing their internal mechanics, symbolic systems stand out for their transparency and intelligibility. For instance, symbolic systems excel not only in making decisions based on established rules and logic but also in illuminating the rationale behind these decisions. This capability for self-explanation stems from the system's proficiency in delineating and expressing the logical sequence from input to outcome, thus offering insight into its decision-making process. Consequently, these systems can uncover underlying principles, decision-making criteria, and even inherent biases in their reasoning, thereby augmenting the transparency and reliability of AI implementations.

In this study, our objective is to utilize stable diffusion coupled with symbolic reasoning to develop a cost-effective solution for translating sensor data into visual representations. We optimize the decoding step through symbolic reasoning to effectively address the challenges posed by varying sensor types within the dataset. This approach allows for a more versatile and efficient translation process, enabling the seamless conversion of diverse sensor inputs into coherent visual outputs.


\section{Methodology}
\label{chap:4}

In this section, we offer an in-depth exposition of GeXSe, beginning with an overview of its design and functionalities. Initially, we delineate the architectural blueprint and core capabilities of GeXSe, setting the stage for a comprehensive analysis. Subsequently, we delve into a rigorous discussion encompassing the theoretical framework underpinning GeXSe, including problem formulation and the application of theoretical principles within the GeXSe system. This approach ensures a holistic understanding of the system's foundation and operational mechanics.

\subsection{GeXSe Overview}



GeXSe, depicted in Figure \ref{overallstructure}, is a generative system designed to interpret diverse sensor data into unified vision domain explanations, utilizing a modular encoder-decoder architecture. This system allows for the independent training of its encoder and decoder components, enhancing its adaptability and efficiency. Despite their separate training process, the encoder and decoder operate within a unified downstream task structure. Within this framework, the encoder is adept at distilling the essential features of diverse sensor data into a universal semantic representation, facilitating an efficient end-to-end learning process that circumvents the need for additional labeling at the inference stage. The decoder, on the other hand, is finely tuned to translate these encoded representations into vision domain narratives, optimized for zero-shot translation to significantly reduce computational requirements. Unlike traditional approaches, GeXSe integrates multiple sensor channels directly into the video output, providing users with a detailed understanding of the contributions of different sensors in the environment, thereby offering a more comprehensive insight into the dynamics of smart spaces.

\subsection{Sensor Data Knowledge Distillation}




The process of cross-domain sensor data fusion tackles the integration of varied sensor datasets from multiple domains, capitalizing on their variety to improve interpretation and prediction across different applications \cite{fan2020buildsensys,zheng2015methodologies}.

To illustrate, let's consider a set of users, denoted as \(\mathcal{U} = \{u_1, \ldots, u_n\}\), where each user \(u\) has a corresponding dataset of raw sensor data represented by \(\mathcal{D}_{u}\). We consolidate these individual datasets into a comprehensive collection, \(\mathcal{D}^\star = \{\mathcal{D}_{u_1}, \ldots, \mathcal{D}_{u_n}\}\) to facilitate a unified analysis. The core objective here is to apply a knowledge distillation strategy, transforming this aggregated sensor data into a more abstract, latent space that captures the underlying patterns and features across all users. This transformation is typically achieved using a teacher-student model framework, designed to harness and transfer knowledge efficiently. 

The employed semantic teaching model is XLNet\cite{yang2019xlnet}, a pre-trained model that leverages permutation-based language modeling. This model efficiently captures bidirectional context, thereby surpassing BERT in sequence order comprehension. Consequently, XLNet facilitates the generation of improved embeddings for nuanced language understanding tasks.

In this framework, the teacher model takes descriptive language annotations as input and generates latent embeddings,  \(\mathcal{Z}_{teacher}\), as output. These embeddings encode the semantic content of the activities described by the annotations, serving as a high-level understanding of the sensor data's underlying patterns and structures. Meanwhile, the student model, which is designed to process sensor data, strives to produce latent embeddings \(\mathcal{Z}_{student}\) that closely approximate \(\mathcal{Z}_{teacher}\).

To facilitate this distillation, we segment each user's sensor dataset \(\mathcal{D}_u\) into non-overlapping, fixed-length temporal windows \(\mathcal{W}_u = \{w_1, \ldots, w_q\}\), each comprising \(z\) seconds of consecutive sensor data. The distillation objective can then be articulated as the optimization problem of refining the sensor data within each window \(w_i\) into a distilled representation \(\ell_i\), such that the collection \(\mathcal{L}_u = \{\ell_1, \ldots, \ell_q\}\) for each user \(u\) aligns with the latent embeddings provided by the teacher model.  Despite the fundamentally different nature of the sensor data inputs compared to textual annotations, this alignment enables the student model to interpret the sensor data in a manner that is semantically consistent with the language model's understanding of the activities.

Given the inherently personalized nature of sensor data within smart spaces, our methodology facilitates the translation of disparate sensor data types into a unified semantic space. This translation serves as a foundational step towards establishing a universal representation within our system, addressing the challenge of sensor data individualization.

\subsection{Multi-Task Learning}
\subsubsection{Multi-Task Learning (MTL) Definition}

In our framework, multi-task learning is operationalized through a model that concurrently learns to output a representation and perform classification, thereby synergizing two objectives. Specifically, the model is designed to optimize the following dual tasks: firstly, to distill raw sensor data into a semantically meaningful representation within a latent space, \(\mathcal{Z}\); and secondly, to accurately classify the type of activity, \(a \in \mathcal{A}\), that generated the sensor data.

Formally, let \(\mathcal{D}\) be the dataset containing raw sensor readings, and \(\mathcal{T} = \{\tau_{repr}, \tau_{class}\}\) represent the set of tasks for representation learning and classification, respectively. The multi-task learning model, \(M\), is a function that maps an input \(x \in \mathcal{D}\) to a tuple \((y_{repr}, y_{class})\), where \(y_{repr} \in \mathcal{Z}\) is the latent representation and \(y_{class} \in \mathcal{A}\) is the predicted activity label:

\[
M: x \mapsto (y_{repr}, y_{class})
\]

\subsubsection{MTL Loss}

The training objective is to minimize a loss function \(L\) that is a composite of the losses for both representation learning, \(L_{repr}\), and classification, \(L_{class}\), effectively enabling the model to learn a shared representation that serves both tasks:

\begin{equation}
L = \alpha L_{\text{repr}}(y_{\text{repr}}, y_{\text{true}}) + \beta L_{\text{class}}(y_{\text{class}}, a_{\text{true}})
\label{MLT_loss}
\end{equation}

where \(\alpha\) and \(\beta\) are the weighting coefficients that balance the importance of each task during training, \(y_{true}\) is the ground truth latent representation, and \(a_{true}\) is the true activity label. This joint optimization encapsulates the essence of multi-task learning in our method, promoting a learning process that is both efficient and robust, particularly suitable for the complex and nuanced domain of smart space sensor data analysis.

\subsubsection{Activity Classification via Cross-Entropy Loss}

For the task of multi-class activity classification within our model, we employ the cross-entropy loss function to evaluate the difference between the predicted probability distributions of the model and the actual distribution of activity labels. Given a finite set of \(k\) possible activities \(\mathcal{A} = \{a_1, \ldots, a_k\}\), the model predicts a probability distribution \(P(x) = (p_1, \ldots, p_k)\) for each input sample \(x\), where each \(p_i\) represents the predicted probability of \(x\) being associated with activity \(a_i\).

The cross-entropy loss \(L_{class}\) for an individual input sample is defined as:

\[
L_{class}(P, a_{true}) = -\sum_{i=1}^{k} y_{i} \log(p_i)
\]

Here, \(y\) is a one-hot encoded vector representing the true activity label, with \(y_i = 1\) if the true activity is \(a_i\) and \(y_j = 0\) for all \(j \neq i\), and \(p_i\) is the model's predicted probability that the input \(x\) belongs to class \(i\).

The objective during the training phase is to minimize the cross-entropy loss \(L_{class}\) over all samples in the training dataset. This minimization drives the adjustment of the model's parameters to align the predicted probability distribution more closely with the true distribution of the labels, thereby enhancing the model's precision in identifying the correct activity class among the multiple categories.

\begin{figure*}
  \centering
  \includegraphics[width=.75\linewidth]{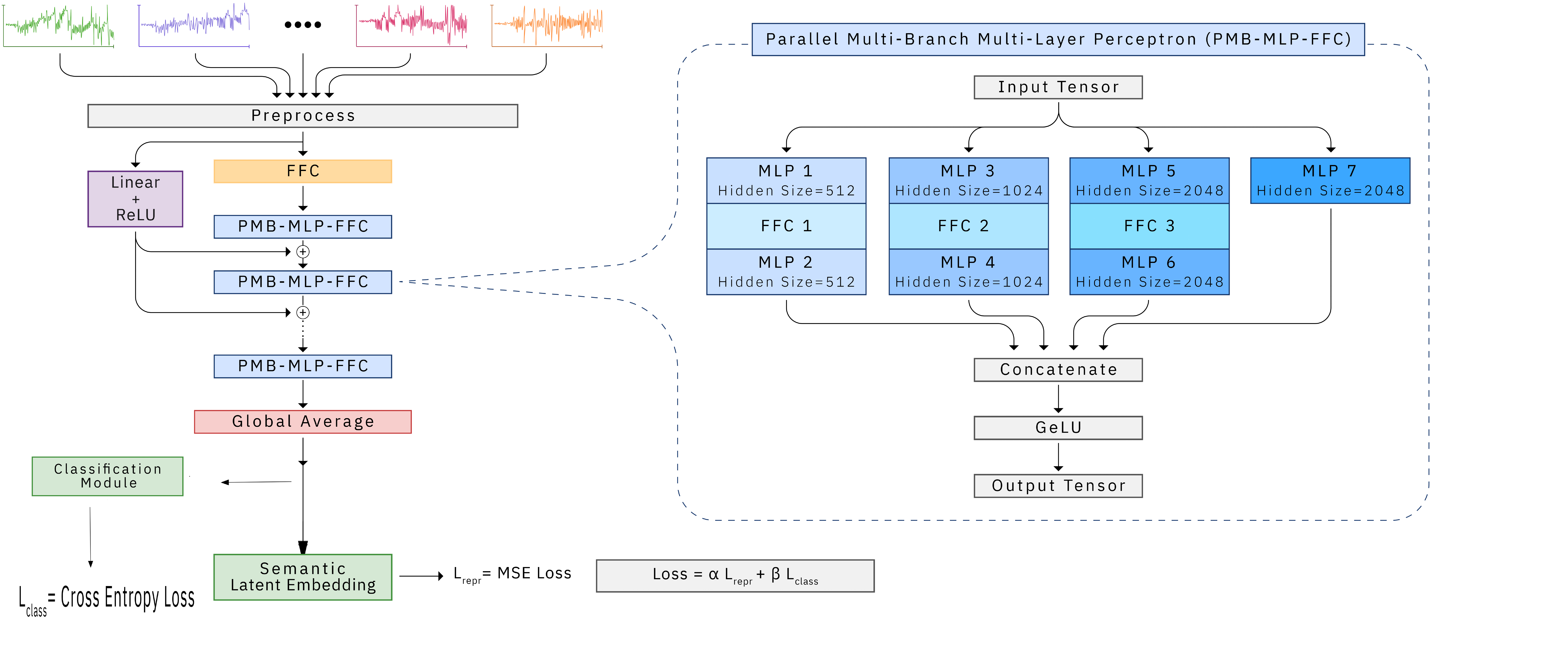}
  \vspace{0 em} 
  \caption{Encoder model (left) incorporating a novel multi-task learning loss to concurrently optimize for classification and semantic representation. It features a unique PMB-MLP-FFC module (right), adeptly extracting comprehensive features from different sensor types.}
  \label{encoder_overallstructure}
\end{figure*}

\subsubsection{Representation Learning via MSE Loss}

To enhance our model's representation learning capability, we employ the Mean Squared Error (MSE) loss function. This approach focuses on minimizing the element-wise squared differences between the predicted latent representation, \(Q(z|x)\), and the target latent representation as ground truth, \(P(z)\), where \(z\) represents the latent embeddings and \(x\) is the input sensor data.

The MSE loss, \(L_{repr}\), for an input \(x\) is defined as:

\[
L_{repr}(P, Q) = \frac{1}{N} \sum_{i=1}^{N} (P(z_i) - Q(z_i|x))^2
\]

Here, \(N\) is the dimensionality of the latent space, \(z_i\) is the \(i\)-th element of the latent vector, \(P(z_i)\) corresponds to the \(i\)-th element of the target latent representation, and \(Q(z_i|x)\) is the \(i\)-th element of the predicted latent representation for the input \(x\). This loss function aims to reduce the discrepancy between the predicted and actual latent representations on an element-wise basis.

Minimizing \(L_{repr}\) during the training phase encourages the model to adjust its parameters such that \(Q(z|x)\) more accurately approximates \(P(z)\) for each element. This is guiding the model towards generating latent representations that are not only meaningful but also semantically aligned with the descriptive nuances conveyed by language. This methodological choice underscores our commitment to achieving a deep semantic synchronization between sensor-derived data and linguistic representations, pivotal for nuanced understanding and analysis within smart environments.

\subsubsection{Addressing Activity Representation Challenges through Multi-Task Learning}

Representing activities in smart environments extends beyond merely mapping the same activities to consistent latent representations. A particular challenge, exemplified by discrepancies in activity labels such as "Open Door 1" versus "Open Door 2" illustrates the need for a model capable of distinguishing between subtly different activities while concurrently merging their representations into a unified latent space. 

\begin{align*}
    &\text{\small Challenge:} \\
    &\quad \textit{\small Differentiate} \, (\text{\small "open door 1"}, 
     \text{\small "open door 2"}) \, \text{\small while} \\
    &\quad \textit{\small Consolidating} \, \text{\small into} \, \mathcal{Z}_{\text{\small universal}}
\end{align*}

This scenario underscores the critical role of multi-task learning within our framework. Our model is designed to navigate the dual objectives of identifying distinct tasks and harmonizing their representations simultaneously. The presence of overlapping activities within sensor datasets—where activities may be identical in some perspectives and divergent in others—mandates a solution that not only categorizes these activities accurately but also aligns them within a universal latent space, denoted as \(\mathcal{Z}_{\text{universal}}\).
\begin{align*}
    &\text{Objective:} \\
    &\quad \textit{Discern} \, (\mathcal{A}_{\text{identical}}, \mathcal{A}_{\text{divergent}}) \, \text{and} \, \textit{Align} \, \text{within} \, \mathcal{Z}_{\text{universal}}
\end{align*}

This dual requirement amplifies the intricacy of our task, compelling the model to not only distinguish between different activities within the same dataset but also to ensure their semantic integration where relevant, as fulfilled by the universal latent space representation, \(\mathcal{Z}_{\text{universal}}\). Leveraging multi-task learning, our model adeptly handles these complex objectives, enhancing activity recognition and representation in smart environments.

 


\subsection{Encoder Model}


Figure~\ref{encoder_overallstructure} (left panel) delineates our encoder's design, trained on three public sensor datasets with 300 epochs and a learning rate of \(1e-3\), utilizing the AdamW optimizer. Upon preprocessing, we pass the data through a Fast Fourier Convolution (FFC) layer for efficient initial feature extraction. The core innovation, the PMB-MLP-FFC module, supports an arbitrary number (\textit{n}) of PMB-MLP-FFC blocks in sequence with \textit{n-1} residual connections and ReLU activation layers to suit various data complexities while avoiding the vanishing gradients problem. In this study, we implemented two instances of the PMB-MLP-FFC module, linked through a single residual connection, facilitating diverse feature extraction through its multi-branch structure with variable hidden layers. Drawing on the strengths of MLP for local feature analysis \cite{vaswani2017attention, lian2021mlp} and FFC for global perspectives \cite{chi2020fast}, our architecture fosters a synergistic local-global feature extraction capabilities—distinctly diverging from conventional inception network frameworks \cite{szegedy2015going}. The architecture comes to a global average pooling layer, which condenses the feature maps into a unified form suitable for subsequent tasks: a classification module powered by MLP and FFC and cross-entropy loss, and a semantic latent embedding module based on a single MLP layer optimized with MSE loss.

\subsubsection{ Fast Fourier Convolution}
\label{enc_ffc_chap}
The Fast Fourier Convolution (FFC), introduced by Chi et al.~\cite{chi2020fast}, represents a novel approach in machine learning, predominantly applied within computer vision. This study extends its application to smart home sensors, aiming to harness FFC for enhanced feature extraction in Human Activity Recognition (HAR). Given the inherent frequency and periodicity in human activities, we capitalized on FFC's potential for robust feature delineation and irrelevant feature filtration. The pseudo-code in Algorithm~\ref{alg:cap} details the FFC process, converting sensor data into the frequency domain for channel-wise analysis. This conversion facilitates one-dimensional convolution, normalization, and activation via the ReLU function, allowing pattern learning in Fourier-transformed data. The uniform treatment of multichannel data in the Fourier domain underscores the versatility of this approach, demonstrating sensor-type agnosticism and signal characteristic discernment.

  


\begin{algorithm}[b]
\caption{Application of Fast Fourier Convolution on Sensor Data \cite{chi2020fast}}\label{alg:cap}
\begin{algorithmic}
\STATE 
\STATE {\textbf{Input Data}: \(x\)}
\STATE \hspace{0.5cm} \(x_{real}, x_{imag} = \text{Real\_FFT}(x)\)
\STATE \hspace{0.5cm} \(X = \text{Concatenate}([x_{real}, x_{imag}], \text{dim}=channel)\)
\STATE \hspace{0.5cm} \(X = \text{ReLU}(\text{BatchNorm}(\text{Conv1d}(X)))\) 
\STATE \hspace{0.5cm} \(x_{real'}, x_{imag'} = \text{Split}(X), \text{dim}=channel)\)
\STATE \hspace{0.5cm} \(x = \text{iFFT}(x_{real'}, x_{imag'})\)
\STATE 
\end{algorithmic}
\end{algorithm}


\subsubsection{PMB-MLP-FFC (Parallel Multi-Branch Multi-Layer Perceptron Fast Fourier Convolution) Module} 
\label{enc_prev_chap}

As elucidated in the previous chapter, the heterogeneity in smart space sensor configurations and data typologies necessitates an architecture adept at accurately interpreting such diverse sensor information. 

\textit{Convolutional Neural Networks} (CNNs) have been extensively validated as effective for analyzing smart space sensor data, as evidenced by recent studies~\cite{arrotta2022dexar,pouyanfar2018survey}. Despite their prevalent adoption, CNNs, which rely on convolution as opposed to the simple matrix multiplications found in \textit{Multi-Layer Perceptrons} (MLPs), introduce additional computational complexity and execution time due to their intricate operational mechanisms~\cite{tolstikhin2021mlp}. Notably, there have been instances where a singular MLP layer outperformed its CNN counterparts in specific tasks~\cite{kaur2021speech}.


In response, we introduce the \textbf{PMB-MLP-FFC} module (illustrated in Figure~\ref{encoder_overallstructure}, right panel), engineered to capture patterns and extract a wide array of feature types across disparate datasets. This module comprises four distinct parallel branches that process an input tensor with unique contributions while preserving the input dimensionality. These branches are differentiated by their hidden sizes.

Three of these branches each stack an MLP expanding the input tensor's dimensionality by a factor of two, an FFC, and a subsequent MLP that acts as a bottleneck, shrinking the tensor's dimension to a quarter of its original size. This design strategy ensures that each branch contributes equally to the output tensor's dimension. The fourth branch, an MLP-only pathway with the same pattern of a fully connected layer followed by a GELU activation function, similarly contracts its output to a quarter of the input size. Following the individual processing, the output tensors from all branches are concatenated, ensuring the final output tensor matches the input tensor's size. A GELU activation function is applied to this concatenated tensor to produce the final output.

\subsection{General Purpose Decoder}

We introduce a framework designed for sensor-based video explanation, leveraging videos pre-trained across various types and latent representations derived from language models. This setup enables the direct translation of sensor data into video descriptions at a rate of 24 frames per second. Our methodology involves a decoder training strategy that aggregates diverse activity videos, pairing them with corresponding language embeddings. This initial pairing serves to train the decoder, which is subsequently employed to interpret latent representations derived from sensor data across various domains. This approach is specifically tailored to scenarios where computational resources are limited, acknowledging the complexity of training vision domain models like 3D UNet in comparison to models processing one-dimensional sensor data. 


\subsubsection{Stable Diffusion for Activity Description}

\begin{figure}[t]
  \centering
  \includegraphics[width=1\linewidth]{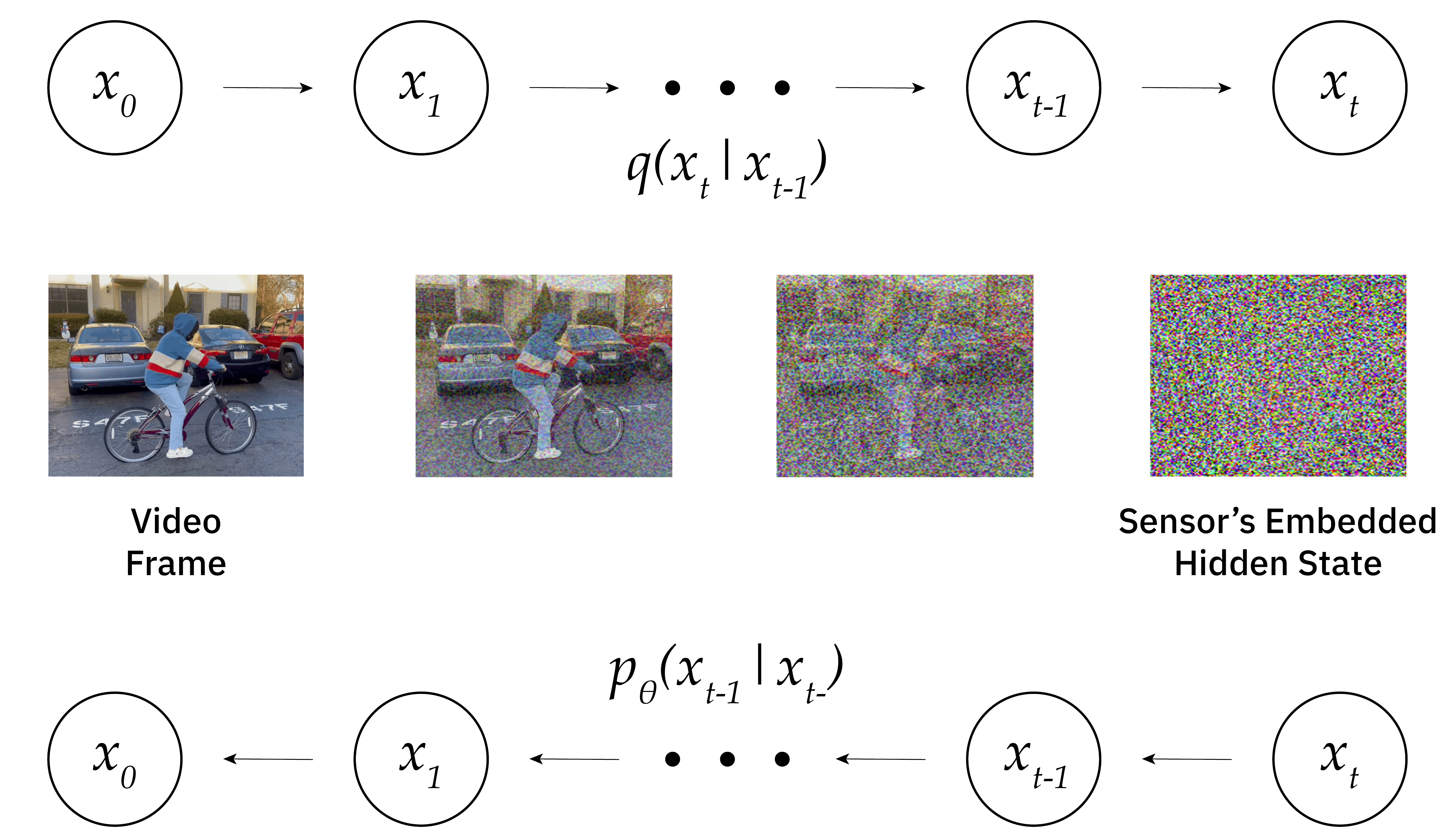}
  \caption{sensor to video generation process}
  \label{diffusionStructure}
\end{figure}

Our proposed approach leverages the concept of stable diffusion, mapping sensor data to visual representations through a Markov chain-based process. The framework, as delineated in Figure ~\ref{diffusionStructure}, introduces a systematic diffusion from the domain of sensor inputs to visual outputs over \(T\) discrete steps. This process, represented by \(q(x_{1:T}|x_0)\), incrementally infuses noise into the data, thereby facilitating the generation of new latent vectors that gradually conform to a target Gaussian distribution characterized by mean \(\mu_t\) and variance \(\sigma_t^2\), with \(\beta\) serving as a control parameter.

\begin{equation}  
q(x_{1:T}|x_0) = \prod_{t=1}^T q(x_t|x_{t-1})
\label{eq:13}
\end{equation}

\begin{equation}  
q(x_t|x_{t-1}) = \mathcal{N}(x_t; \mu_t = \sqrt{1-\beta_t}x_{t-1}, \sigma_t^2 = \beta_t I)
\label{eq:14}
\end{equation}

Crucially, our generative model adapts to the inherent variability of the sensor data via a conditional diffusion model. Equation~\ref{eq:18} formalizes this adaptive mechanism, where \(y\) represents specific activity conditions detected by the sensors, such as a person running. The encoder first converts these conditions into a latent embedding, which the decoder then utilizes to synthesize a corresponding visual narrative.

\begin{equation}  
p_{\theta}(x_{0:T}|y) = p_{\theta}(x_T) \prod_{t=1}^T p_{\theta}(x_{t-1}| x_t, y)
\label{eq:18}
\end{equation}

\begin{equation}  
\nabla_{x_t} \log p_{\theta}(x_t|y) = \nabla_{x_t} \log p_{\theta}(x_t) + s \cdot \nabla_{x_t} \log p_{\theta}(y|x_t)
\label{eq:19}
\end{equation}

\begin{equation}  
\hat{\mu}(x_t|y) = \mu_{\theta}(x_t|y) + s \cdot \sigma_{\theta}(x_t|y) \nabla_{x_t} \log f_{\phi}(y|x_t, t)
\label{eq:20}
\end{equation}

This framework not only simplifies the sensor-to-visual domain translation but also enriches the generated video with nuanced details reflective of the underlying sensor data, offering a sophisticated means to visualize sensor-detected activities.

\textbf{Practical Application Example:} As depicted in Figure 1, consider a scenario where sensor arrays in a smart home environment detect the pattern of someone closing a dishwasher. The encoder translates this specific pattern into a latent vector, capturing the dishwasher closing activity's unique characteristics. Subsequently, the decoder, informed by this latent vector and leveraging the stable diffusion process, generates a short video segment that visually depicts the dishwasher closing activity. This showcases the model's ability to transform abstract sensor data into a visually understandable representation, thereby offering a more comprehensive insight into the dynamics of smart spaces.

\subsection{Neural Symbolic Reasoning }

  
  

\begin{figure*}[t]
  \centering
  \includegraphics[width=1\linewidth]{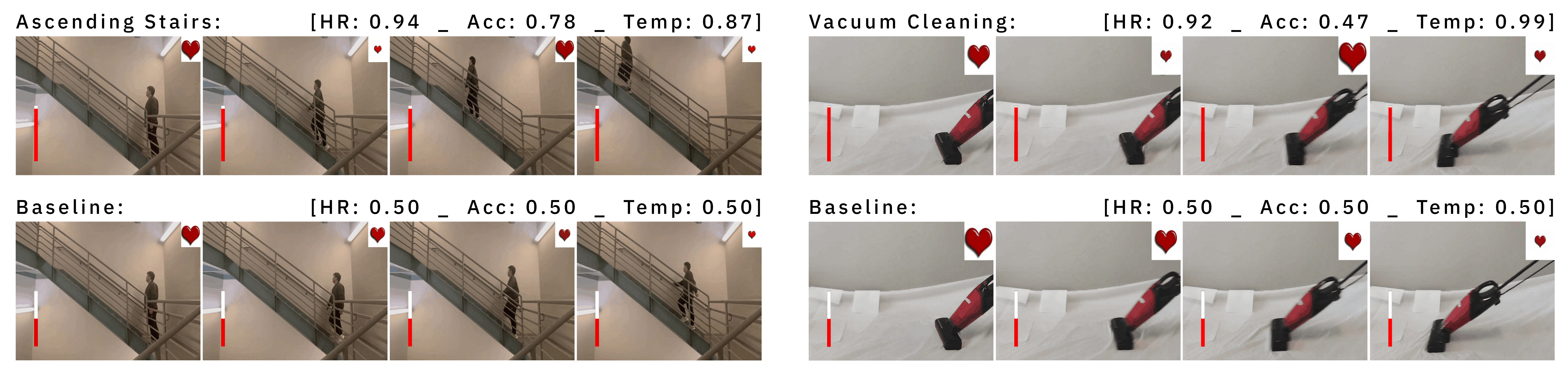}
  \vspace{0 em} 
  \caption{Four frames of the symbolic reasoning output for "Ascending Stairs" and "Vacuum Cleaning" activities, illustrating variations in sensor activation levels per activity in comparison to their respective baselines. Normalized sensor activation values for heart rate, accelerometer, and temperature are noted. The overall output frames are 24.}
  \label{fig:combined_result_output}
\end{figure*}


Beyond standard video narratives, real-world applications often encounter variations in sensor domain descriptions for identical activities performed by different individuals or in different conditions. By leveraging insights from sensor contributions quantified through sensor channel activations derived from the encoder, we integrate symbolic reasoning to enable nuanced representation of sensor variations corresponding to the same video descriptions \cite{garcez2019neural}. By establishing this low-cost, zero-shot translation system, our approach enables the generation of contextually rich video descriptions adaptable to diverse sensor inputs without necessitating further decoder retraining to accommodate such information.

This process involves the following steps:

\begin{enumerate}
    \item \textbf{Quantification of Sensor Activation:} Sensor data, encompassing a diverse array of physical phenomena, is quantified by the model to produce activation levels for each sensor channel, symbolically represented by a continuous value in the interval \([0, 1]\).
    \item \textbf{Symbolic Encoding:} These quantified activations are then symbolically encoded into the video output. Symbolic reasoning is employed to map specific activation levels to predetermined visual cues or modifications within the video, such as altering the speed of video or changing temperature indicator based on the activation of the accelerometer and temperature sensor, respectively.
\end{enumerate}

Formally, let \(S = \{s_1, s_2, \ldots, s_n\}\) denote the set of sensor channels, and \(A = \{a_1, a_2, \ldots, a_n\}\) represent the corresponding activation levels, where \(a_i \in [0, 1]\) for each sensor channel \(s_i\). The symbolic reasoning process, denoted as \(f: A \rightarrow V\), maps these activation levels to a set of visual modifications \(V\), thus transforming quantitative sensor data into qualitative visual representations. This mapping is defined by a set of rules or functions that dictate how activation levels influence specific aspects of the video output, enhancing the interpretability and informativeness of the visualized data.

\begin{equation}
    V_i = f(a_i) \quad \forall i \in \{1, 2, \ldots, n\}
\end{equation}

This problem definition underscores our approach to leveraging symbolic reasoning for the intuitive and dynamic representation of sensor data within video outputs, aiming to provide users with a richer understanding of the underlying sensor contributions while obviating the need for retraining the decoder model to offer such information.

Figure~\ref{fig:combined_result_output} illustrates selected outcomes of the symbolic reasoning for two activities, each with two examples of different activation levels, showcasing the variance in heart rate (indicated by the pulse rate of the heart shape in the top right corner), temperature (represented by a red bar in the bottom left corner), and accelerometer readings (reflected through the manipulated frame rate of video) following symbolic reasoning.




\section{Experiments and Results}
\label{chap:5}

In this section, we detail our experimental setup and outcomes, focusing on evaluating the performance of our encoder and the resultant decoder outputs. Initially, we describe the datasets utilized in our study. Subsequently, we assess the encoder's effectiveness in capturing and representing sensor data. Finally, we appraise the quality of our vision domain explanations through human evaluation across various dimensions.

\subsection{Datasets}
In selecting datasets for our experiments, we encountered several challenges that needed to be addressed. The primary issues were the translation of the same activities across different sensor types (using body-mounted sensors, smartphone sensors, and ambient sensors in different datasets) and distinguishing between subtly different activities. To tackle these challenges, we selected three publicly available datasets: PAMAP2, UCI-HAR, and Opportunity. The activity types present in the UCI-HAR dataset can be considered a subset of those in PAMAP2, which helps address the issue of translating the same activity across various sensors. The Opportunity dataset, known for its focus on smart space Human Activity Recognition (HAR), addresses the challenge of classifying similar activities differently, such as distinguishing between "Open Door 1" and "Open Door 2". 

The training and testing datasets are constructed from non-overlapping subjects to ensure generalizability; individuals featured in the testing set were not included in the training phase. The effectiveness of the classification is evaluated using the macro F1 score, serving as the primary metric for testing performance. Below are the details of each dataset:

\subsubsection{PAMAP2}

The PAMAP2\cite{reiss2012introducing} dataset is a rich repository of physical activity data, captured from a group of 9 individuals (comprising 8 males and 1 female), who were instructed to perform a comprehensive array of 18 lifestyle activities. These activities span from daily household chores, such as lying down, sitting, standing, walking, running, cycling, Nordic walking, ironing, vacuum cleaning, jumping rope, as well as ascending and descending stairs, to various leisure pursuits including watching television, working on a computer, driving, folding laundry, house cleaning, and playing soccer. Participants were outfitted with three inertial measurement units (IMUs) attached to their hand, chest, and ankle, which meticulously recorded data across multiple parameters: accelerometer, gyroscope, magnetometer, temperature, and heart rate, over the duration of 10 hours. The constructed PAMAP2 dataset comprises 36 dimensions, with the data segmented into fixed-width sliding windows of 2.56 seconds, each overlapping by 50\%, resulting in 256 readings for every window due to a sampling rate of 100 Hz. These 36 distinct signals formed the input for our model.

\subsubsection{UCI-HAR}

The UCI Human Activity Recognition (HAR) dataset\cite{anguita2013public},  collects activity data from 30 participants using waist-mounted smartphones with inertial sensors, aiming to classify six basic activities: three static postures (standing, sitting, lying) and three dynamic movements (walking, walking downstairs, and walking upstairs). Triaxial linear acceleration and angular velocity data were recorded at a 50 Hz sampling rate, segmented into 2.56-second sliding windows with a 50\% overlap, resulting in 128 readings per window. Preprocessing included noise reduction via median and third-order low-pass Butterworth filters with a cutoff frequency of 20 Hz, which also served to segregate the acceleration signal into body motion and gravitational components. This meticulous preparation yielded a dataset with 9 processed signal channels, ready for input into deep learning models for comprehensive activity analysis.

\subsubsection{Opportunity}

The Opportunity\cite{roggen2010collecting} activity recognition dataset is a robust collection of naturalistic activities, acquired in a sensor-rich environment leveraging 72 different environmental and body sensors across 12 subjects, 4 subjects were released to the public. Utilizing 15 networked sensor systems spanning 10 sensor modalities, integrated within the environment, embedded in objects, and worn on the body, the dataset is particularly advantageous for benchmarking a variety of activity recognition methodologies. For our analysis, we focused solely on data from the inertial measurement units (IMUs), including triaxial accelerometers, gyroscopes, and magnetometers, among others, while omitting quaternion data. This approach provided us with 77 distinct signal channels. The dataset was sampled at 30 Hz, and we employed 3-second windows for data extraction, which yielded 90 samples per window, forming the input for our model.

\subsection{Encoder Performance}

This section is dedicated to assessing the classification capabilities of our encoder. Experimental results demonstrate that our design not only facilitates the development of a generalized encoder but also enhances classification performance in comparison to models dedicated solely to classification tasks.

We conducted a comprehensive comparative analysis of various smart space Human Activity Recognition (HAR) models, highlighting differences in architecture and parameters for a broad evaluation. Table \ref{tab:5} delineates the models ranging from basic configurations such as CNNs and LSTM networks to more complex models like AROMA and DeepConvLSTM, which have a significantly larger parameter count. Our model features an FFC module for global feature representation, prompting us to include a comparison with a typical transformer-based model (THAT) known for its global feature-capturing capabilities through attention mechanisms. Additionally, we benchmark against UniTS, a Fourier-based machine learning model, which, akin to THAT, aligns with our model in terms of parameter count, enabling a direct comparison to elucidate differences in performance and efficiency. Brief descriptions of these models are provided below:

\begin{itemize}
    \item vLSTM\cite{mutegeki2020cnn}: The architecture comprises multiple LSTM cells arranged in a stack, intended for analyzing features within the sensor dataset. This recurrent configuration facilitates the recognition of temporal dynamics and spatial relationships. The performance of this recurrent model in discerning patterns within time series sensor data is demonstrated.
    \item CNN-HAR\cite{yang2015deep}: The architecture primarily consists of a CNN utilized for recognizing human activities based on data from multiple sensors. To integrate feature maps from diverse sensors, the model formulates layers that extract pertinent features essential for distinguishing various activities. The objective is to synergize feature learning and classification, thereby enhancing the overall recognition process.
    \item AROMA\cite{peng2018aroma}: It constitutes a multitask learning paradigm employing a residual-based architecture. It simultaneously integrates convolutional, pooling, and fully connected layers as hidden components for multitask learning.
    \item DeepConvLSTM\cite{ordonez2016deep}: It represents a multimodal wearable sensor model founded on LSTM and CNN architectures. This model incorporates a recurrent module designed to capture temporal dynamics and facilitate seamless sensor fusion. In contrast to CNN alone, the inclusion of the recurrent module enables the effective modeling of time series data's temporal variations.
    \item THAT\cite{li2021two}:  It introduces a novel approach to human activity recognition (HAR) by exploiting WiFi signals' unique variations due to human movement. It addresses the limitations of existing WiFi-based HAR methods by capturing both time-over-channel and channel-over-time features using a two-stream structure. The model incorporates multi-scale convolution augmented transformers to detect range-based patterns, demonstrating superior performance and efficiency across real experiment datasets.
    \item UniTS\cite{li2021units}: It integrates Short-Time Fourier Transform (STFT) principles into deep neural networks for sensory time series classification, eliminating the need for domain-specific feature engineering or intensive hyper-parameter tuning. By making STFT weights trainable within its architecture, UniTS adeptly combines time and frequency domain information to discern discriminative patterns in sensory data, demonstrating its effectiveness across various datasets with a notable improvement in classification performance.
\end{itemize}



\begin{table}[t]
  \centering 
  \caption{Benchmark HAR models for comparative analysis}
  \label{tab:5}  
  \begin{tabular}{c|c}
    \toprule
    Model Name & Number of Parameters \\
    \midrule
    vLSTM\cite{mutegeki2020cnn} &  120,397 \\
    CNN-HAR\cite{yang2015deep} &  310,069 \\
    AROMA\cite{peng2018aroma} & 515,552  \\
    DeepConvLSTM\cite{ordonez2016deep} & 3,444,205  \\
    THAT\cite{li2021two} &  6,837,502 \\
    UniTS\cite{li2021units} &  6,536,475  \\
    GeXSe (Ours)  &  6,392,693 \\
    \bottomrule
  \end{tabular}
\end{table}

As illustrated in Table \ref{tab:combined}, our encoder consistently outperforms the aforementioned models in terms of precision, recall, and F1 scores across different datasets. Specifically, in the UCIHAR dataset, which features only 6 activities, our model outperforms all others by achieving the highest classification accuracy, as well as its own best performance, with an F1 score of 0.96. For the PAMAP2 dataset, despite the challenge of classifying a larger variety of activities (12 in total), GeXSe still achieves a notable performance with the F1 score of 0.89, indicating its better consistency in performance across diverse activity sets compared to competitors. Lastly, for the Opportunity dataset, known for its imbalance issues, GeXSe again sets the benchmark with the highest F1 score of 0.83. This demonstrates its relative robustness and effectiveness in dealing with imbalanced datasets, where it significantly outperforms other models in recognizing activities despite the challenging data conditions.

The lower performance in datasets with a higher number of activities and/or imbalance, like PAMAP2 and Opportunity, suggests that traditional transformers might not be the optimal choice for handling sensor data with these characteristics. Our encoder, with its FFC-MLP design, is specifically engineered to leverage both global and local features effectively, addressing potential issues commonly associated with transformers. This design choice is pivotal in enhancing our model's adaptability and performance across datasets with varying characteristics, demonstrating its robustness and the effectiveness of incorporating both global and local feature analysis in activity recognition tasks.


\begin{table*}[ht]
  \centering
  \caption{Comprehensive comparison of different HAR models across PAMAP2, UCIHAR, and Opportunity datasets.}
  \label{tab:combined}
  \begin{tabular}{c|ccc|ccc|ccc}
    \toprule
    & \multicolumn{3}{c|}{PAMAP2 Dataset} & \multicolumn{3}{c|}{UCIHAR Dataset} & \multicolumn{3}{c}{Opportunity Dataset} \\
    Model Name & Precision & Recall & F1 & Precision & Recall & F1 & Precision & Recall & F1 \\
    \midrule
    vLSTM\cite{mutegeki2020cnn} & 0.86 & 0.86 & 0.85 & 0.92 & 0.92 & 0.92 & 0.70 & 0.66 & 0.67 \\
    CNN-HAR\cite{yang2015deep}  & 0.87 & 0.87 & 0.87 & 0.95 & \textbf{0.95} & 0.95 & 0.73 & 0.71 & 0.71 \\
    AROMA\cite{peng2018aroma} & 0.87 & 0.87 & 0.87 & 0.94 & 0.94 & 0.94 & 0.72 & 0.72 & 0.72 \\
    DeepConvLSTM\cite{ordonez2016deep}  & 0.86 & 0.87 & 0.87 & 0.92 & 0.92 & 0.92 & 0.74 & 0.73 & 0.73 \\
    THAT\cite{li2021two} & 0.87 & 0.86 & 0.87 & 0.87 & 0.87 & 0.87  & 0.72 & 0.72 & 0.72 \\
    UniTS\cite{li2021units} & 0.87 & 0.89 & 0.89 & 0.92 & 0.92 & 0.92 & 0.79 & 0.78 & 0.78 \\
    GeXSe (Ours) & \textbf{0.89} & \textbf{0.90} & \textbf{0.89} & \textbf{0.96} & \textbf{0.95} & \textbf{0.96} & \textbf{0.84} & \textbf{0.83} & \textbf{0.83} \\
    \bottomrule
  \end{tabular}
\end{table*}

Examining more closely, the PAMAP2 dataset exhibits robust performance across activities, though standing and descending stairs reveal lower recognition rates, as shown in Table~\ref{tab:2}. The UCIHAR dataset, referenced in Table~\ref{tab:4}, achieves higher recognition rates, albeit with difficulty distinguishing between walking and walking upstairs, indicating the closeness of these activities in the latent embedding space. The Opportunity dataset, mentioned in Table~\ref{tab:6}, highlights a high recognition rate for drinking activities due to a predominance of such data, leading to imbalance. This imbalance impacts intra-class classification, such as differentiating between opening drawer 1 and drawer 2.

\begin{figure*}
  \centering
  \includegraphics[width=\linewidth]{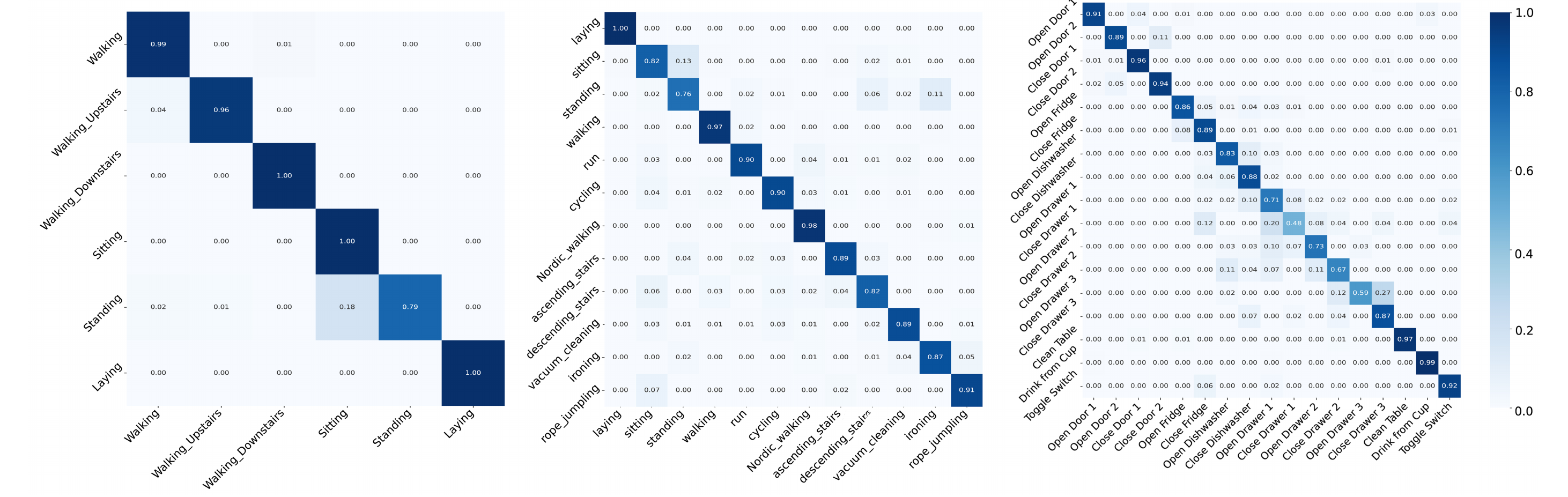}
  \caption{Normalized Confusion Matrices: UCIHAR (left), PAMAP2 (middle), Opportunity (right)}
  \label{confusion}
\end{figure*}

The detailed insights from the normalized confusion matrices in Figure \ref{confusion} reveal varying intra-class classification performance levels across datasets. The UCIHAR dataset showcases the highest performance while The PAMAP2 dataset highlights some challenges in differentiating between a few activities like standing and sitting, indicating the closeness of such activities in the latent embedding space. Meanwhile, the Opportunity dataset illustrates more often difficulties in intra-class classification, underlining the model's struggle with distinguishing closely related activities. Nevertheless, our model's architecture supports intra-class classification more effectively than the competitors, demonstrating its capability to produce consistent embedded vectors for classification and decoding purposes.

    


\begin{table}[tp]
  \centering
    \caption{ Activity analysis of PAMAP2 dataset }
    \label{tab:2}
    \begin{tabular}{c|c|c|c}
    \toprule
    Activity Name &Precision &Recall &  F1-Score \\
    \midrule
    
Lying & 0.99 & 1.00 & 1.00 \\
Sitting & 0.84 & 0.82 & 0.83 \\
Standing & 0.76 & 0.76 & 0.76 \\
Walking & 0.97 & 0.97 & 0.97 \\
Running & 0.93 & 0.90 & 0.92 \\
Cycling & 0.93 & 0.90 & 0.92 \\
Nordic walking & 0.90 & 0.98 & 0.94 \\
Ascending stairs & 0.92 & 0.89 & 0.90 \\
Descending stairs & 0.75 & 0.82 & 0.78 \\
Vacuum cleaning & 0.90 & 0.89 & 0.89 \\
Ironing & 0.91 & 0.87 & 0.89 \\
Rope jumping & 0.76 & 0.91 & 0.83 \\

      \bottomrule
   \end{tabular}
\end{table}

    


\begin{table}[tp]
  \centering
    \caption{ Activity analysis of UCIHAR dataset }
    \label{tab:4}
    \begin{tabular}{c|c|c|c}
    \toprule
    Activity Name &Precision &Recall &  F1 \\
\midrule
Walking & 0.95 & 0.99 & 0.97 \\ 
Walking Upstairs & 0.99 & 0.96 & 0.97 \\ 
Walking Downstairs & 0.99 & 1.00 & 0.99 \\ 
Sitting & 0.83 & 1.00 & 0.91 \\ 
Standing & 1.00 & 0.79 & 0.88 \\ 
Laying & 1.00 & 1.00 & 1.00 \\
\bottomrule
   \end{tabular}
\end{table}


\begin{table}[h]
  \centering
    \caption{ Activity analysis of Opportunity dataset }
    \label{tab:6}
    \begin{tabular}{c|c|c|c}
    \toprule
Action & Precision & Recall & F1-Score \\ 
\midrule
Open Door 1 & 0.95 & 0.91 & 0.93 \\ 
Open Door 2 & 0.94 & 0.89 & 0.91 \\ 
Close Door 1 & 0.95 & 0.96 & 0.95 \\ 
Close Door 2 & 0.90 & 0.94 & 0.92 \\ 
Open Fridge & 0.89 & 0.86 & 0.88 \\ 
Close Fridge & 0.81 & 0.89 & 0.85 \\ 
Open Dishwasher & 0.83 & 0.83 & 0.83 \\ 
Close Dishwasher & 0.69 & 0.88 & 0.77 \\ 
Open Drawer 1 & 0.67 & 0.71 & 0.69 \\ 
Close Drawer 1 & 0.60 & 0.48 & 0.53 \\ 
Open Drawer 2 & 0.79 & 0.73 & 0.76 \\ 
Close Drawer 2 & 0.62 & 0.67 & 0.64 \\ 
Open Drawer 3 & 0.97 & 0.59 & 0.73 \\ 
Close Drawer 3 & 0.73 & 0.87 & 0.79 \\ 
Clean Table & 1.00 & 0.97 & 0.98 \\ 
Drink from Cup & 0.99 & 0.99 & 0.99 \\ 
Toggle Switch & 0.94 & 0.92 & 0.93 \\
\bottomrule
   \end{tabular}
\end{table}





\subsection{Decoder Performance}

Model interpretation and multimodal data representation are two distinct concepts within the field of data science and machine learning, each addressing different aspects of working with complex data. Model interpretation focuses on understanding and explaining the decisions, predictions, or outputs generated by a machine learning model. It aims to make the model's operations transparent, allowing humans to comprehend how input data is transformed into outputs, identify the significance of different features, and assess the model's reliability and biases. This is crucial for trust, accountability, and ethical considerations, especially in applications affecting real-world decisions.

On the other hand, multimodal data representation concerns the integration and processing of data from multiple sources or types, such as text, images, audio, and video, within a single model or analytical framework. The challenge here lies in effectively combining diverse data formats to leverage their combined predictive power or insights, which requires sophisticated data preprocessing, feature engineering, and model architecture adjustments. Multimodal representation is essential in applications like automatic content recognition, sentiment analysis where input comes in various forms, and enhancing the model's understanding by providing it with a richer, more comprehensive view of the data.

\begin{figure}[t]
  \centering
  \hbox{\hspace{-1.5 em} \includegraphics[width=1\linewidth,height=0.2\textheight]{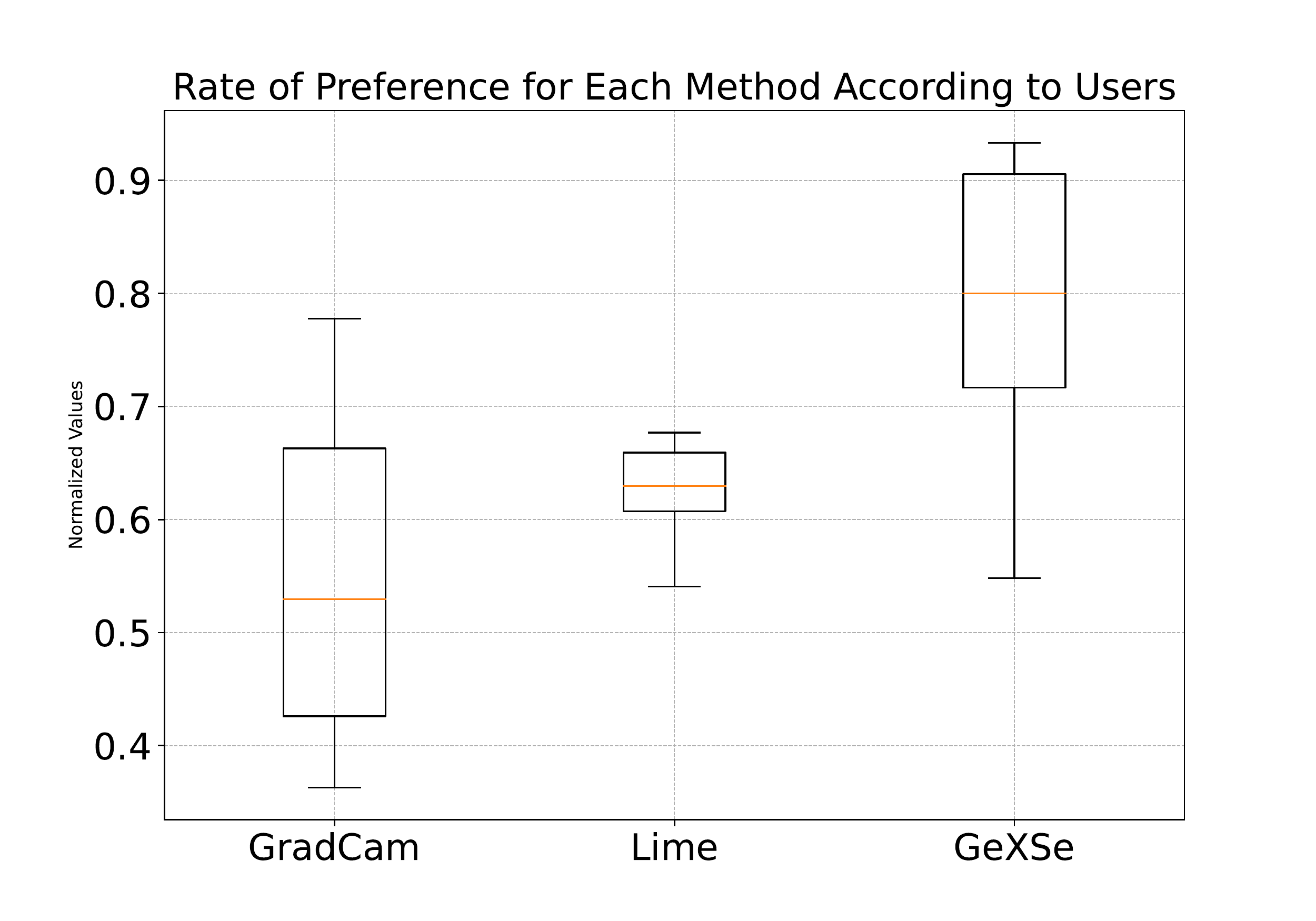}}
  \vspace{0 em} 
  \caption{Human Assessment of Method Preferences }
  \label{human_eval_pref}
\end{figure}

\begin{figure}[t]
  \centering
  \hbox{\hspace{-1.5 em} \includegraphics[width=1\linewidth,height=0.2\textheight]{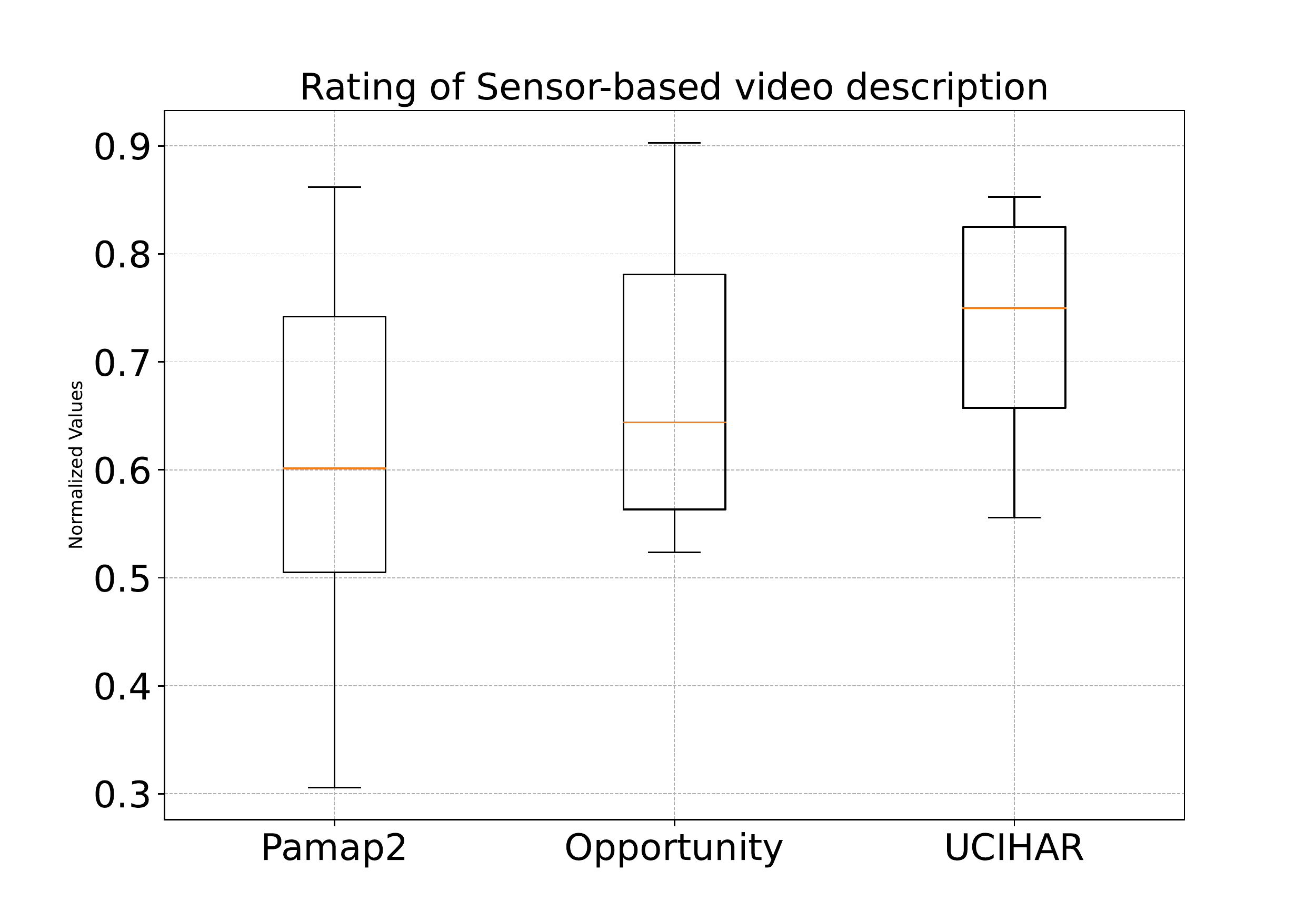}}
  \vspace{0 em} 
  \caption{Human Assessment of Sensor-Based Video Descriptions }
  \label{human_eval_sensor_video}
\end{figure}

\subsubsection{Sensor Channel Activation Feedback Survey}

Our approach employs a multimodal methodology, indicating that the conventional techniques for assessing model interpretability are not applicable to our method. In light of this, we rely on human evaluation to assess the performance of our model, which aims to produce videos where variations in input data rates are perceptibly reflected in the output videos. Hence, we designed two surveys to conduct such human evaluation studies. 

In the first survey, we focus on assessing human observers' preferences for different sensor data representation methods within video generation models. This survey aims to compare the effectiveness of our proprietary method against two established model explanation techniques, Gradcam and LIME, in communicating the relevance of sensor data to the generated representation.

In the Representation Preference Evaluation Survey, participants are presented with a series of representations of activities, each annotated with visual highlights generated by different explanation techniques. These highlights aim to indicate areas of the representation that are most influenced by the sensor data inputs. The participants' task is to review these representations and rate how distinguishable and interpretable the activities are represented (from "Strongly Disagree" to "Strongly Agree") in each method. This choice is reflective of the participants' subjective perception of which technique provides the clearest and most understandable visual explanation.

The second survey serves to validate whether changes in acceleration, heartbeat rate, and temperature input data rates are clearly conveyed and discernible in the model-generated videos from a human observer perspective, thus examining the perceived visual influence of the input data on the rendered videos. The premise is that an increase in an input data rate (e.g. acceleration) should result in a perceptible increase in the corresponding visualized measure (e.g. higher frame rate). This principle applies similarly to other data types. 

In the Sensor Channel Activation Feedback Survey, participants are presented with a series of videos rendered by our model using varying input rates, along with a baseline video for comparison.  Their task is to assess the agreement level (from "Strongly Disagree" to "Strongly Agree") that the variations in input rates are perceptibly reflected in the visualized changes between videos. This allows us to evaluate how understandably our model communicates the relationship between data inputs and video outputs to external viewers.

The results presented below are derived from surveys conducted both in-person and online. We gathered feedback from a total of 60 participants with different backgrounds. To mitigate any potential bias introduced by the order of the survey, we divided participants into separate groups, each comprising 30 individuals, to assess distinct topics. The total valid response is about 900.

\subsubsection{Representation Preference Evaluation Survey}


Figure ~\ref{human_eval_pref} illustrates the results of the Representation Preference Evaluation Survey. The survey aims to ascertain the preferred methodology for feature extraction from sensor data models. The normalized preference ratings for three distinct methodologies—Gradcam, LIME, and our proposed GeXSe technique—are presented in box plot format.

From the data, it is evident that GeXSe secures higher median preference ratings in comparison to Gradcam and LIME, indicating a marked preference among users for this method. The interquartile range and the whiskers of the box plots reflect the variance in user preferences, with the occurrence of outliers further depicting the diversity of opinions. Notably, the LIME methodology demonstrates a more concentrated distribution of ratings, suggesting a consensus in user perception regarding its interpretability. Conversely, the wider spread in the ratings for GeXSe and Grad-CAM suggests a broader range of user preferences, yet with an inclination towards higher favorability.


Figure ~\ref{human_eval_sensor_video} presents the results from the Sensor Channel Activation Feedback Survey. The survey is designed to evaluate the perceived effectiveness of video content modifications triggered by sensor channel activations. Ratings are normalized on a scale from 0 to 1 and presented in box plot format.

For the PAMAP2 dataset, user consensus is quantified regarding the visual representation of sensor data: the pulse rate depicted by a heart shape for the heartbeat sensor, the level of a red bar for temperature data, and the video frame rate corresponding to accelerometer readings. The Opportunity and UCIHAR datasets, however, focus on representing accelerometer data via frame rate alterations alone. UCIHAR, with its narrower scope of activity types, achieves a higher median rating, suggesting that the representation's simplicity correlates with stronger user consensus. The same pattern is closely followed in the Opportunity dataset case. In contrast, PAMAP2 shows a more dispersed range of user responses, attributed to its variety of sensor types and activity types. This diversity suggests that an increase in the types of sensor and activity data correlates with a wider range of user interpretations and assessments.

\section{Limitations}
\label{chap:6}

Beyond the contributions of our study, we aim to outline several limitations pertinent to our research.

\begin{itemize}
    \item Firstly, the model's capability is currently confined to recognizing a restricted set of activities, which does not fully capture the complexity of real-life scenarios where multiple activities may occur simultaneously. This limitation suggests that the single-activity detection approach may not be sufficient for more intricate situations.
    \item Secondly, the inference time for the stable diffusion model, with our settings tuned to $t=500$, is substantial. To enhance the model's feasibility for real-time or near-real-time applications, it is imperative to reduce this inference duration.
    \item Thirdly, the diffusion process at this stage can occasionally produce unstable or inaccurate outputs. This imperfection necessitates attention as the model may render frames that do not precisely represent the input descriptions.
    \item Lastly, the majority of the datasets used to train and evaluate smart space models are derived from indoor settings. Such datasets lack heterogeneity and complexity present in outdoor or less structured environments. To bridge the gap, there is a pressing need for diverse datasets that more accurately capture the broad spectrum of human activities in various settings.

\end{itemize}

\section{Conclusion and Future Work}
\label{chap:7}


In this study, we introduced GeXSe, a generative multi-task framework for Human Activity Recognition (HAR) in smart spaces, which proved efficient in fusing sensor data for not only classification but also for generating explanatory visual narratives, accessible to users without expertise in machine learning. By training our encoder to distill cross-domain sensor inputs into a unified semantic representation and leveraging a pre-trained vision domain decoder alongside symbolic reasoning, we effectively manage the variability inherent in sensor data. Our experiments across several public datasets demonstrate the encoder's ability to provide a universal activity representation and achieve superior classification F1 scores, with human evaluations further affirming the preferability of our methodology. Future work will explore extending this methodology to more complex real-world scenarios, aiming for faster, practical applications to enhance sensor data comprehension.

\section{Acknowledgments}

This work was supported by the National Science Foundation (NSF) as part of the Center for Smart Streetscapes, under NSF Cooperative Agreement EEC-2133516. DataCity Smart Mobility Testing Ground is jointly funded by Middlesex County Resolution 21-821-R, New Jersey Department of Transportation and Federal Highway Administration Research Project 21-60168.

\bibliography{sample-base}
\bibliographystyle{plain}

\end{document}